\documentclass[superscriptaddress,twocolumn]{revtex4}

\usepackage{amsfonts}
\usepackage{amsmath}    
\usepackage[normalem]{ulem}
\usepackage{bm}
\usepackage{graphicx}

\newcommand{\ket}[1]{\vert#1\rangle}

\def\opone{\leavevmode\hbox{\small1\kern-3.8pt\normalsize1}}
\usepackage{color}

\begin{document}

\title{Quantum storage of entangled telecom-wavelength photons in an erbium-doped optical fibre}

\author{Erhan~Saglamyurek}
\affiliation{Institute for Quantum Science and Technology, and Department of Physics \& Astronomy, University of Calgary, 2500 University Drive NW, Calgary, Alberta T2N 1N4, Canada}
\author{Jeongwan~Jin}
\affiliation{Institute for Quantum Science and Technology, and Department of Physics \& Astronomy, University of Calgary, 2500 University Drive NW, Calgary, Alberta T2N 1N4, Canada}
\author{Varun~B.~Verma}
\affiliation{National Institute of Standards and Technology, Boulder, Colorado 80305, USA}
\author{Matthew~D.~Shaw}
\affiliation{Jet Propulsion Laboratory, California Institute of Technology, 4800 Oak Grove Drive, Pasadena, California 91109, USA}
\author{Francesco~Marsili}
\affiliation{Jet Propulsion Laboratory, California Institute of Technology, 4800 Oak Grove Drive, Pasadena, California 91109, USA}
\author{Sae~Woo~Nam}
\affiliation{National Institute of Standards and Technology, Boulder, Colorado 80305, USA}
\author{Daniel~Oblak}
\affiliation{Institute for Quantum Science and Technology, and Department of Physics \& Astronomy, University of Calgary, 2500 University Drive NW, Calgary, Alberta T2N 1N4, Canada}
\author{Wolfgang~Tittel}
\affiliation{Institute for Quantum Science and Technology, and Department of Physics \& Astronomy, University of Calgary, 2500 University Drive NW, Calgary, Alberta T2N 1N4, Canada}

\maketitle

\textbf{The realization of a future quantum Internet requires processing and storing quantum information at local nodes, and interconnecting distant nodes using free-space and fibre-optic links\cite{Kimble_08}. Quantum memories for light\cite{Lvovsky_09} are key elements of such quantum networks. However, to date, neither an atomic quantum memory for non-classical states of light operating at a wavelength compatible with standard telecom fibre infrastructure, nor a fibre-based implementation of a quantum memory has been reported. Here we demonstrate the storage and faithful recall of the state of a 1532 nm wavelength photon, entangled with a 795 nm photon, in an ensemble of cryogenically cooled erbium ions doped into a 20 meter-long silicate fibre using a photon-echo quantum memory protocol. Despite its currently limited efficiency and storage time, our broadband light-matter interface brings fibre-based quantum networks one step closer to reality. Furthermore, it facilitates novel tests of light-matter interaction and collective atomic effects in unconventional materials.} 
 
The end of the last century witnessed the invention of, and important steps towards, several paradigm-shifting applications of quantum information science, including computers with unprecedented computational power\cite{Ladd_10}, unbreakable secret key distribution\cite{Gisin_02}, and measurement devices having ultimate precision\cite{Giovannetti_11}. Combining these applications in the so-called quantum Internet\cite{Kimble_08} requires transmitting quantum states encoded into photons between, and storage of quantum states in, nodes of the network. While the quantum Internet can leverage existing telecom fibre networks, standard (classical) repeater technology cannot be used to build large-scale networks, due to a fundamental restriction of quantum mechanics known as the no-cloning theorem\cite{Gisin_02}. Hence, classical repeaters, generally comprised of erbium-doped fibre amplifiers, need to be replaced with quantum repeaters, which include pairs of entangled photons, entanglement swapping, and light-matter interfaces that allow storing and manipulating quantum states of light\cite{Sangouard_11}.  

Despite enormous success in developing suitable light-matter interfaces during the past decade (for recent reviews see [\citenum{Lvovsky_09, Sangouard_11,Bussieres_13}]), a memory for non-classical states of light encoded into telecom-wavelength (i.e. approximately 1550 nm) photons --- the most natural choice for a quantum network --- still remains to be demonstrated. Considering the  most popular materials  -- alkaline atoms (in particular caesium and rubidium), and rare-earth-ion doped crystals -- the reasons for this challenge are twofold: First, Cs and Rb lack easily accessible atomic transitions, i.e. transitions starting at an electronic ground state, at around 1550 nm wavelength.  
Second, erbium (a rare-earth element and the seemingly obvious choice due to its telecom-wavelength transition and extensive use in fibre amplifiers) has so-far eluded all attempts to store non-classical states of light with a fidelity above the classical limit, albeit important steps towards this goal have recently been reported\cite{Lauritzen_10,Dajczgewand_13}. Furthermore, in addition to telecom-wavelength storage, there is another milestone that, if accomplished, would significantly benefit fibre-based quantum communication networks: the storage of non-classical states by means of light-atom interaction in optical fibres. This promises a simplified and robust setup -- comparable to the use of  an erbium fibre amplifier in standard telecom networks. We note that the use of a hollow core photonic crystal fibre filled with caesium atoms is a promising step in this direction\cite{Spraque_14}.  

Here we demonstrate that the apparent limitations of erbium for optical quantum memory can be overcome, and, furthermore, that quantum states of light can indeed be stored in impurities doped into an optical fibre. More precisely, employing a commercially available erbium-doped fibre cooled to around 1 K and exposed to a suitably chosen magnetic field, we report the storage and faithful recall of 1532 nm photons that are entangled with 795 nm photons using the atomic frequency comb (AFC) quantum memory protocol\cite{Riedmatten_08, Afzelius_09}.  

\begin{figure*}
\begin{center}
\includegraphics[width=\textwidth,angle=0]{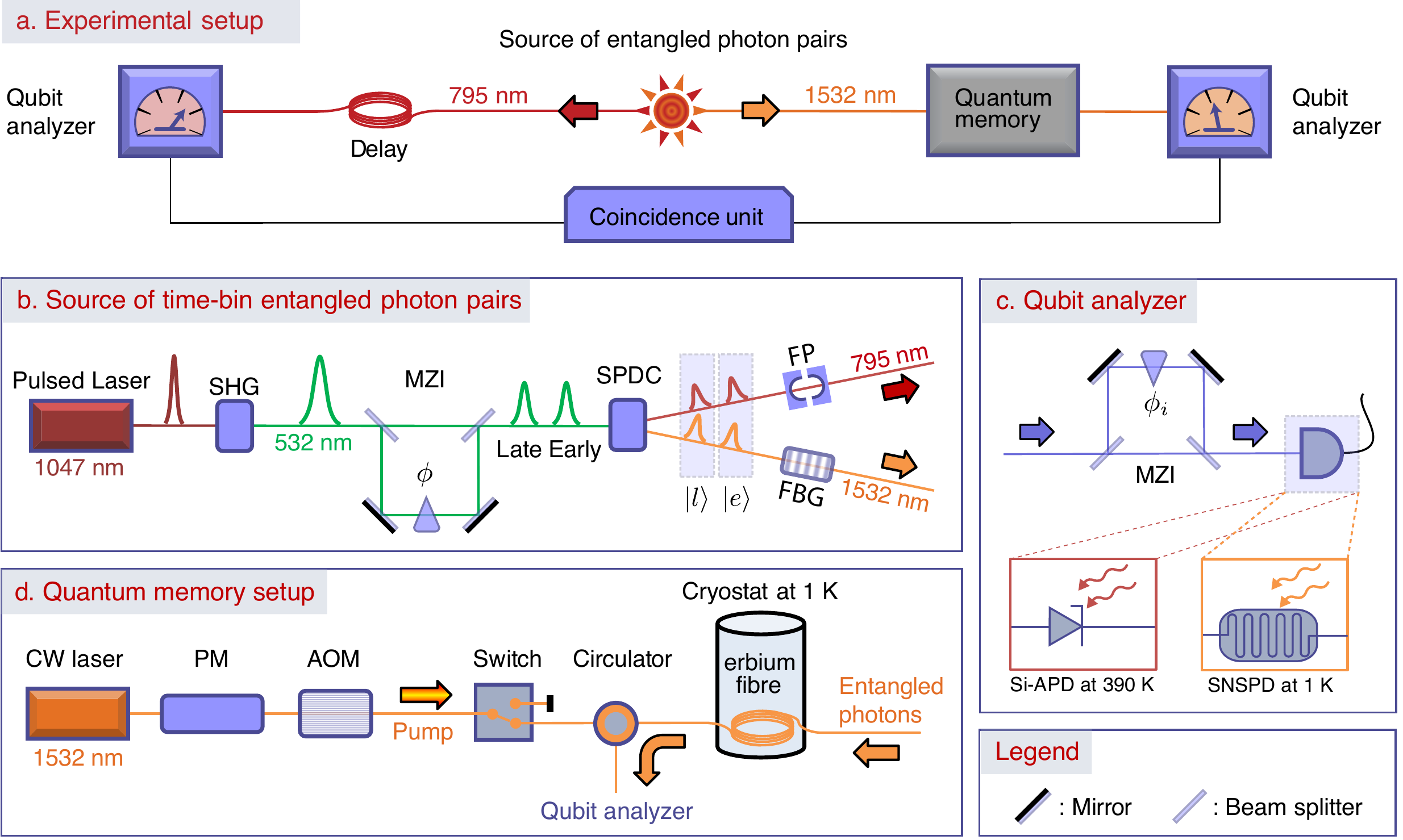}
\caption{\textbf{Experimental setup.} \textbf{a) Schematic.} Our setup consists of three parts: the source of entangled photon pairs, the quantum memory, and analyzers (including a coincidence unit) that allow projecting the joint photon state onto qubit product states. Depending on the measurement, the entangled photons are either directly sent towards the analyzers, or the 1532 nm photon is first stored for 5 ns in an Erbium-doped fibre.
 \textbf{b) Photon pair source.} SHG -- second harmonic generation; MZI -- Mach-Zender interferometer; SPDC -- spontaneous parametric down conversion; FP -- Fabry-Perot cavity; FBG -- fiber Bragg grating.
\textbf{c) Analyzers.} SNSPD - superconducting nano-wire single photon detectors. 
 \textbf{d) Quantum memory.} PM -- phase modulator; AOM -- acousto-optic modulator. For details on all components see the Methods section.}
\label{setup}
\end{center}
\end{figure*}

Our experimental setup, sketched and further explained in Fig.~\ref{setup} and in the Methods section, is composed of three parts: a source of time-bin entangled photon pairs, a quantum memory for photons, and analyzers that allow projection measurements with the members of each entangled photon pair. First, using spontaneous parametric down-conversion of short laser pulses, we generate photon pairs with members at 795 nm and 1532 nm wavelength in a time-bin entangled bi-photon state given by
\begin{equation}\label{timebinentangled}
\ket{\phi^{+}}=\frac{1}{\sqrt{2}}\left(\ket{e,e}+\ket{l,l}\right).
\end{equation}
\noindent Here $\ket{i,j}$ denotes a quantum state in which the 795-nm photon has been created in the temporal mode $i$, and the 1532-nm photon has been created in the temporal mode $j$. Furthermore, $i,j\in[e,l]$, and $\ket{e}$ and $\ket{l}$ label early and late temporal modes, respectively. The spectra of the 795 nm photon and the telecom photon are filtered  to 6 GHz and 10 GHz, respectively, to allow subsequent storage of the 1532 nm photon in the erbium-doped fibre. 
\begin{figure*}
\begin{center}
\includegraphics[width=1\textwidth,angle=0]{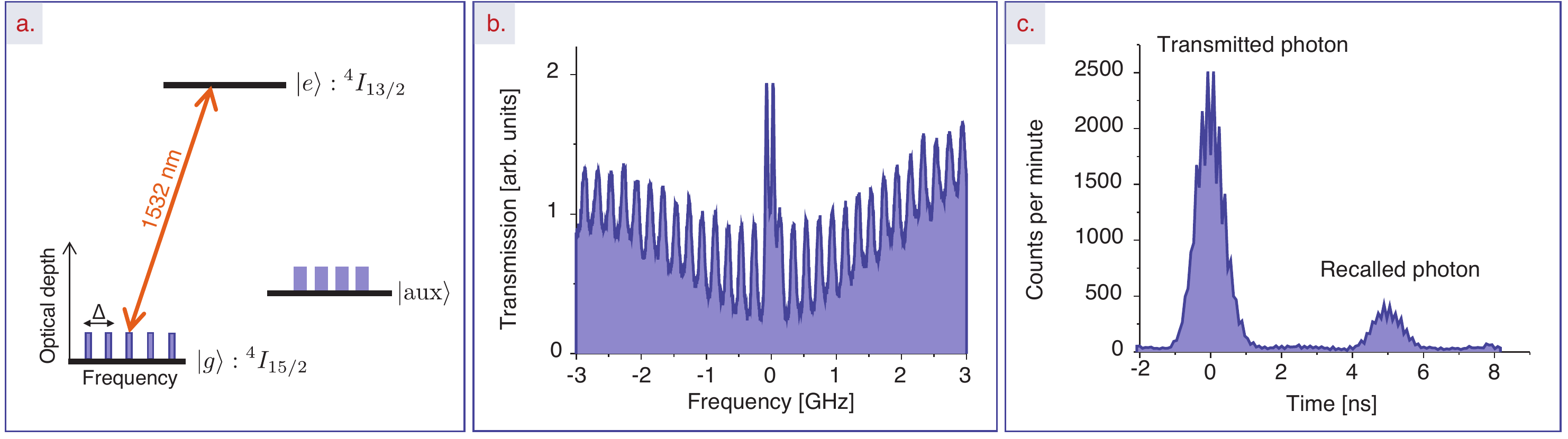}
\caption{\textbf{Storage of telecom-wavelength photons in a broadband AFC memory.} \textbf{a) Simplified erbium level scheme.} To generate AFCs, we frequency-selectively transfer population from the $^4I_{15/2}$ electronic ground state ($\ket{\mathrm{g}}$) through the $^4I_{13/2}$ excited state ($\ket{\mathrm{e}}$) into an auxiliary state ($\ket{\mathrm{aux}}$).
\textbf{b) Typical atomic frequency comb.} The total AFC bandwidth is 8 GHz (only 6 GHz are shown), and the peak spacing 200 MHz, leading to 5 ns-long storage. The variation of the AFC's envelope is an artifact. \textbf{c) Photon storage.} 1532 nm wavelength photons in a single temporal mode are stored and retrieved after 5~ns with $\sim$1\% efficiency (for more details see the Supplementary Information), and detected by an SNSPD. Also shown are detections due to directly transmitted (i.e. not absorbed) photons. The detection-time jitter of the detector used for this measurement is around 800 ps, which restricts one from observing the true duration of the original and re-emitted photons (around 44 and 55~ps, respectively).}
\label{AFC}
\end{center}
\end{figure*} 
Second, to store the 1532 nm photons, we use a cryogenically-cooled, commercially available erbium-doped fibre (see the Supplementary Information) in conjunction with the atomic frequency comb (AFC) quantum memory protocol\cite{Riedmatten_08, Afzelius_09}. AFC-based storage relies on tailoring an inhomogeneously broadened atomic transition (in our case the $^4I_{15/2}\leftrightarrow^4I_{13/2}$ transition in Er$^{3+}$-ions) into a series of absorption lines that are equally spaced by frequency $\Delta$ -- that is the atomic frequency comb (see Figure~\ref{AFC}a, b). The absorption of a photon then creates a collective atomic excitation described by:
\begin{equation} \label{dicke1}
\ket{\Psi}_A=\frac{1}{\sqrt{N}}\sum_{j=1}^{N} c_{j} e^{i2\pi \delta_j t} e^{-ikz_j} \ket{g_{1} ,\cdots e_{j} ,\cdots g_{N}}  
\end{equation}
\noindent where $N$ is the number of atoms in the ensemble and $\delta_j$ is the transition frequency of the $j^{th}$ atom with respect to the input photon's carrier frequency, i.e. the detuning. $z_j$ and $c_j$ denote the position of the $j^{th}$ atom within the medium and the excitation probability amplitude for each atom, respectively. Following the creation of the collective atomic excitation, the different components in Eq.~\ref{dicke1} start accumulating different phases owing to different detunings $\delta_j$ of the excited ions. However, due to the discrete  and periodic nature of the possible atomic transition frequencies, $\delta_j=m\,\Delta$ ($m \in\mathbb{Z}$), all phases automatically line up at time $\tau=1/\Delta$, resulting in the re-emission of the photon in its original quantum state (see Fig. \ref{AFC}c). 
Feed-forward control of the mapping between an input and an output mode, e.g. on-demand recall in the temporal domain or frequency selective recall in the spectral domain, can be achieved with the application of demonstrated techniques\cite{Afzelius_10b,Sinclair_13}.
Furthermore, under certain circumstances, the retrieval process can approach unit efficiency\cite{Afzelius_09,Afzelius_10}. The AFC protocol, implemented in rare-earth-ion doped crystals, has already shown great promise as a quantum memory for quantum information processing applications, including a retrieval fidelity exceeding  99.5\%\cite{Zhou_12}, an efficiency of 56\%\cite{Sabooni_13}, more than 5 GHz bandwidth\cite{Saglamyurek_11}, the possibility to store a large number of temporal and spectral modes\cite{Usmani_10, Bonarota_11, Sinclair_13}, recall on demand\cite{Afzelius_10b} and feed-forward control\cite{Sinclair_13}, waveform preserving storage\cite{Jin_13}, and the possibilities to store members of entangled photon pairs\cite{Saglamyurek_11, Clausen_11} plus teleport photonic quantum states into crystals\cite{Bussieres_14}.
	
Finally, to analyze the joint state of the photon pairs before and after storage of the 1532 nm photon, we perform individual projection measurements onto certain time-bin qubit states 
 \begin{equation}
 \label{projection}
 \ket{\psi}=\alpha\ket{e}+\beta e^{i\theta}\ket{l},\hspace{1cm}\alpha^2+\beta^2=1.
 \end{equation}  
The  measurements are henceforward also referred to as projections onto eigenstates of the Pauli operators $\sigma_x$ (corresponding to projections onto $(\ket{e}\pm\ket{l})/\sqrt{2}$), $\sigma_y$ (corresponding to projections onto $(\ket{e}\pm i\ket{l})/\sqrt{2}$), $\sigma_z$ (corresponding to projections onto $\ket{e}$ and $\ket{l}$), and superpositions of $\sigma_x$, $\sigma_y$ and $\sigma_z$.   
 
\begin{figure}[ht!]
\begin{center}
\includegraphics[width=\columnwidth,angle=0]{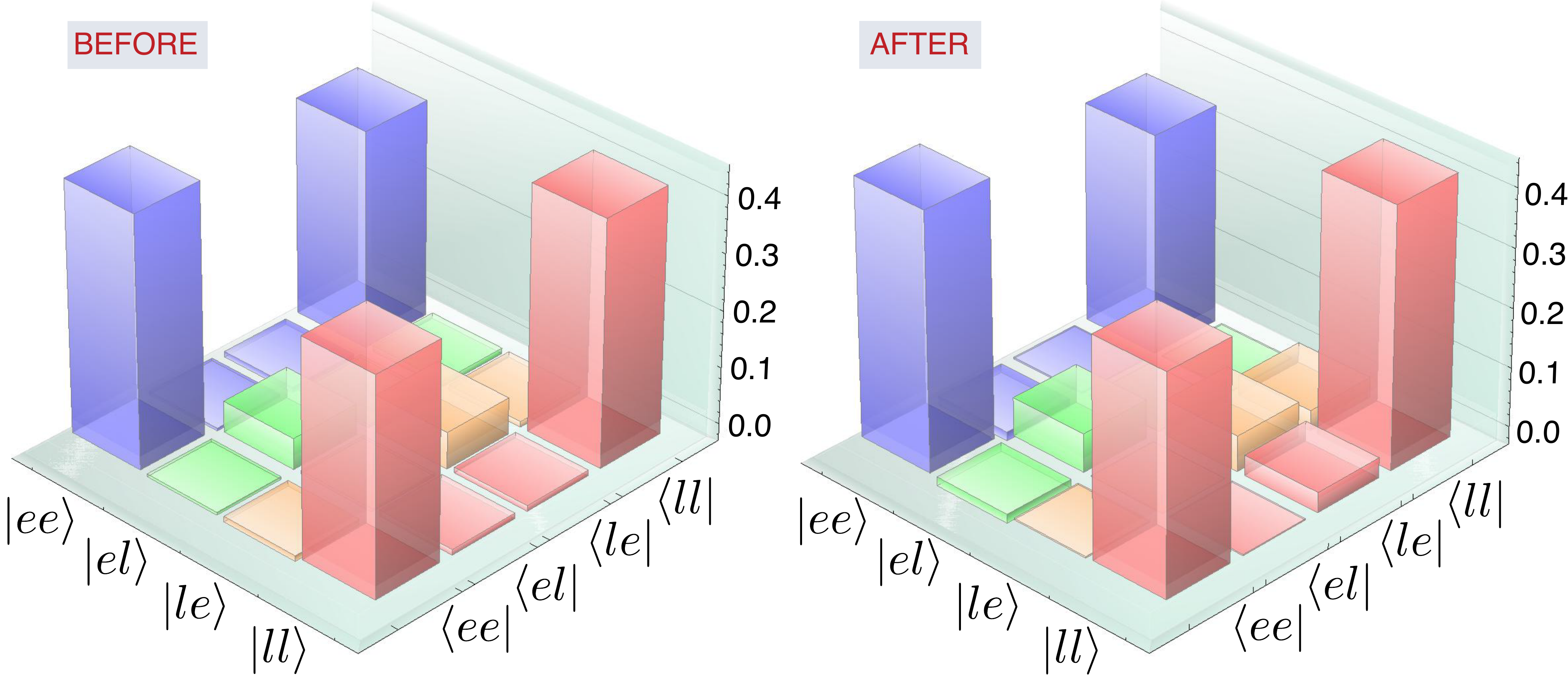}
\caption{\textbf{Reconstructed density matrices.} The figure shows the measured density matrices of the photon pairs before (in) and after (out) storage of the telecom photon. Only the real components are shown -- the absolute values of all imaginary components are below 0.025.}
\label{matrix}
\end{center}
\end{figure}

To experimentally demonstrate our memory's ability to preserve entanglement, we perform a set of projection measurements (see the Methods section) that allows reconstructing the quantum state of the photon pairs before storage in terms of its density matrix using a maximum likelihood method\cite{Altepeter_05}. We then repeat the same measurements after having stored and retrieved the 1532 nm photon. The resulting matrices are shown in Figure~\ref{matrix}, and Table \ref{results} lists various parameters that quantify relevant properties of the two-photon system for each case.  
 
The first parameter of interest is the fidelity, which quantifies the overlap between two quantum states. We find that the fidelities of the photon pair states before and after storage with respect to the maximally entangled $\ket{\phi^{+}}$-state given in Eq. \ref{timebinentangled} are 0.825$\pm$0.004 and 0.808 $\pm$0.048, respectively. Furthermore, we calculate the purity of the photon pair states before and after storage, yielding 0.694 $\pm$ 0.007 and 0.673 $\pm$ 0.047, respectively. The deviation from the optimum value of one in all cases (i.e. concerning the fidelity and the purity) is due to SPDC not creating individual photon pairs, but rather a distribution over even numbers of photons. However, we note that the two measured fidelities, as well as both purities, are equal to within the statistical uncertainty, suggesting that the state has not been altered during storage. To verify this conjecture, we also calculate the input-output fidelity of the quantum state after storage with respect to the state before storage. We find $F_\mathrm{in/out}$= 0.971 $\pm$0.049, confirming that the state did indeed not change.  

Another important parameter is the entanglement of formation $E_F$, which we use to quantify the amount of entanglement in our photon pairs. (Values for $E_F$ range from zero for a separable state to one for a maximally entangled state.) We find $E_F$=0.531 $\pm$0.011 before storage and, similarly, 0.499 $\pm$0.105 after storage. This reflects that the state does not change during storage, and furthermore shows that it is --- and remains --- entangled. 

\begin{table}[t!]
\begin{center}
\scalebox {0.8}{
\begin{tabular}{l c c}
  \hline                        
	\textbf{Quantity} & \centering{\textbf{Before storage}} & \textbf{After storage} \\
	\hline
Fidelity with $\ket{\phi^{+}}$ ($\%$) &82.5$\pm$0.4&80.8 $\pm$4.8\\
Purity ($\%$) & 69.4 $\pm$ 0.7& 67.3 $\pm$4.7\\
Input/Output fidelity ($\%$) & \multicolumn{2}{c}{97.1 $\pm$4.9} \\
Entanglement of formation ($\%$)& 53.1 $\pm$1.1& 49.9 $\pm$10.5 \\
Expected $S_\mathrm{th}$ & 2.39 $\pm$0.01& 2.35 $\pm$0.10 \\
Measured $S$& 2.38 $\pm$0.05& 2.33 $\pm$0.22 \\
  \hline
\end{tabular}
}
\caption{\textbf{Characterization of the two-photon state.} Using reconstructed and ideal density matrices, we calculate the fidelity of our two-photon state before and after storage with the maximally entangled $\ket{\phi^{+}}$-state, its purity, the fidelity of $\rho_\mathrm{out}$ with respect to $\rho_\mathrm{in}$, and the entanglement of formation. The table also shows expected and experimentally obtained values for tests of the CHSH Bell inequality. Uncertainties (one standard deviation) are estimated from Poissonian detection statistics using Monte Carlo simulation. For more details see the Supplementary Information.}
\label{results}
\end{center}
\end{table}

In addition to the tomographic reconstruction, we also perform a Clauser--Horne--Shimony--Holt (CHSH) Bell-inequality test\cite{Clauser_69} before and after storage. This test reveals whether or not the correlations between outcomes of measurements on a bi-partite system can be explained by local realistic theories (LRT). Assuming local realistic theories,  
CHSH derived that
\begin{equation} \label{eq:CHSH}
S=|E(\pmb{a},\pmb{b})-E(\pmb{a},\pmb{b^\prime})+E(\pmb{a^\prime},\pmb{b})+E(\pmb{a^\prime},\pmb{b^\prime})|
\end{equation}
has an upper bound of two:
\begin{equation}
S_{\mathrm{LRT}}\leq 2 \ ,
\end{equation}
while quantum mechanics predicts a maximum value of $S^\mathrm{max}_{\mathrm{QM}}=2\sqrt{2}\approx 2.82$.
Here, $E(\pmb{x,y})$ is the correlation coefficient measured when projecting particle one (i.e. the 795 nm photon) onto  $\ket{x}$, and particle two (i.e. the 1532 nm photon) onto $\ket{y}$. For more details see the Methods section and the Supplementary Information. Before storage, we find $S_\mathrm{in}$= 2.38 $\pm$0.05, and, crucially, $S_\mathrm{out}$= 2.33 $\pm$0.22 after storage (see Table~\ref{results}). Hence, we find a violation of the maximum value allowed by local realistic theories by 7.5 and 1.5 standard deviations, respectively.  
    
Our results show that photon-photon entanglement can be reversibly mapped onto entanglement between a photon and a collective atomic excitation that is delocalized over $\sim$10$^{13}$ erbium atoms spread over twenty meters. Being based on the same material as a classical erbium fibre amplifier, our memory is the first to store non-classical states of light at telecom wavelength, and, furthermore, it is the first quantum memory that employs light-atom interaction in an optical fibre. 
We note that the ease of integration into fibre networks, and its large time-bandwidth product and multimode storage capacity make our light-matter interface attractive for  quantum information processing tasks in addition to storage\cite{Saglamyurek_14}. Furthermore, erbium-based memories may allow interfacing telecommunication photons with superconducting qubits\cite{O'Brien_14}. 

From a more fundamental point of view, as the length of our atomic medium exceeds the pulse length of the light that the absorbers collectively interact with, our results open the path towards a new kind of cavity QED-type experiments\cite{Scully_09}. In addition, they raise the interesting question of whether there is an upper limit to the separation between atomic absorbers --- e.g. the light's wavelength\cite{Dicke_54} --- after which the entanglement inherent in Eq. \ref{dicke1} breaks down. This decoherence would manifest itself as an unusual decrease of storage efficiency.  While our demonstration falls short to answer this question (the average distance was around 10 nm, i.e. $\sim\lambda/100$), the possibility to use impurity-doped optical fibres opens the path towards kilometre-long, very weakly doped memory materials with large inter-dopand separation, i.e. the possibility for an experimental test.
We anticipate that our investigation will benefit the realization of quantum networks, and also trigger more fundamental research towards improved understanding of light-matter interaction in glassy hosts, and collective atomic effects in unconventional materials.

\begin{center}
\textbf{METHODS}
\end{center}
\subsection{Components}

\noindent
\textbf{a) Photon pair source} (see also Fig. \ref{setup}b). A mode-locked laser emits, with 80 MHz repetition rate, 6 ps-long pulses at 1047 nm wavelength. The pules are frequency doubled by means of second harmonic generation (SHG) in a periodically poled lithium niobate (PPLN) crystal, and are directed to an imbalanced Mach-Zender interferometer (MZI) that splits each pulse into two, separated by 1.4 ns.  Spontaneous parametric down conversion in a PPLN crystal yields time-bin entangled photon pairs with members centered at 795 nm and 1532 nm wavelength. Remaining 523 nm light is removed through filters (not shown), the photons from each pair are separated using a dichroic mirror, and their bandwidths are reduced to approximately 6 GHz and 10 GHz  by means of a Fabry-Perot cavity  (for the 795 nm photon), and a fibre Bragg grating (for the 1532 nm photon), respectively, to allow  storage of the 1532 nm photon in the erbium-doped fibre. In addition, the filtering creates entangled pairs that are suitable for quantum teleportation, and, furthermore, would allow storing the 795 nm photon in our Tm:LiNbO$_3$ quantum memory\cite{Saglamyurek_11}.

\noindent
\textbf{b) Analyzers} (see also Fig. \ref{setup}c). The quantum states of the 795 and 1532 nm photons are projected onto time-bin qubit states spanned by two, 1.4 ns-separated temporal modes: we employ either MZIs featuring path-lengths differences of 1.4 ns followed by single photon detectors (SPDs) (allowing projections onto $\ket{\varphi}=(\ket{e} +e^{i\varphi}\ket{l})/\sqrt{2}$), or a delay line followed by an SPD (allowing projections onto $\ket{e}$ and $\ket{l}$, not shown). The SPDs for the 795 nm photons are based on silicon avalanche photo diodes and feature detection efficiencies around 60\% and dark counts around 100 Hz, and the 1532 nm photons are detected using superconducting nano-wire single photon detectors (SNSPDs) with the system efficiency of around 60\% and dark counts around 10 Hz [\citenum{Marsili_13}] (see the Supplementary Information for more details). A coincidence unit and a PC allow assessing the rates with which certain combination of projections occur, e.g. the rate with which the 795 nm photon is projected onto $\ket{e}$ and the 1532 nm photon is projected onto $(\ket{e}+\ket{l})/\sqrt{2}$. 

\noindent
\textbf{c) Quantum memory} (see also Fig. \ref{setup}d). The light from an extended-cavity continuous wave (CW) laser at 1532 nm wavelength is frequency and intensity modulated using a phase modulator and acousto-optic modulator (AOM), respectively. After passing a switch and a circulator, it is sent into the erbium-doped fibre, which is exposed to a $\sim$600 Gauss magnetic field to split erbium energy levels into magnetic sub levels (see Fig. \ref{AFC}a). Erbium-light interaction then leads to frequency-selective persistent spectral hole burning with hole life times on the order of tens of seconds (more information will be published elsewhere\cite{Saglamyurek_14b}) and, after repetition of the burning sequence during 400 ms, to an 8 GHz wide atomic frequency comb (see Fig. \ref{AFC}b; for additional details see the Supplementary Information). After spectral tailoring, we wait 300 ms before storing and retrieving entangled photons -- this is necessary to allow excited atoms to decay, i.e. ensure that recalled photons are not masked by spontaneously emitted photons. To remove the light used for spectral hole burning during waiting and photon storage (700 ms during each experimental cycle), the position of the switch is toggled.

\subsection{Measurements} 
\textbf{a) Density matrices.} To analyze the bi-photon states, we perform joint-projection measurements with the 795 nm and the 1532 nm photons onto time-bin qubit states characterized by $\pmb{a}$, $\pmb{b}$, respectively, where $\pmb{a,b}$ $\in$ [$\pm\sigma_x, \pm \sigma_y,\pm\sigma_z$]. This is done by means of suitably adjusted qubit analyzers, and by counting the number $C(\pmb{a,b})$ of detected photon pairs per time. Recall that, for time-bin quits, projecting a photon's state onto $\pm\sigma_z$ corresponds to detecting the photon in an early or late time bin, respectively, while projections onto $\pm \sigma_x$, $\pm\sigma_y$ and any linear combinations of these correspond to detecting the photon in a specific output of a widely unbalanced  interferometer with appropriately set phase. The phases of the three interferometers (one in the photon pair source plus one per analyzer) are individually locked using a frequency-stabilized laser (not shown in Fig.\ref{setup}) and adjusted using a procedure described previously\cite{Saglamyurek_11}.

From two joint projection measurement, we calculate the normalized joint-detection probability
\begin{equation} \label{detection probability} 
P(\pmb{a,b})=\frac{C(\pmb{a,b})}{C(\pmb{a,b})+C(\pmb{a,-b})}  
\end{equation} 
where $\pmb{b}$ and $\pmb{- b}$ refer to projections onto orthogonal states.
The values of nine different joint-detection probabilities (stemming from all combinations of $\pmb{a,b}\in [\sigma_x,\sigma_y,\sigma_z]$) allow constructing the density matrices for our bi-partite quantum system\cite{Altepeter_05} (see Figure~\ref{matrix}).

\textbf{b) Bell inequality.} To test the CHSH-Bell-inequality, we perform four sets of measurements, each consisting of four joint measurements with projections onto any combination of $\pm\pmb{a}$ (measured on one particle) and $\pm\pmb{b}$ (measured on the other particle), respectively, where $\pmb{a}\otimes\pmb{b}\in[\sigma_y\otimes(\sigma_x+\sigma_y),\sigma_y\otimes(\sigma_x-\sigma_y),\sigma_x\otimes(\sigma_x+\sigma_y),\sigma_x\otimes(\sigma_x-\sigma_y)]$ (the chosen settings allow violating the CHSH-Bell inequality maximally). For each set, we calculate the correlation coefficient  

\begin{equation} \label{eq:corr}
E(\pmb{a},\pmb{b})=\frac{C(\pmb{a},\pmb{b})-C(\pmb{a},\pmb{-b})-C(\pmb{-a},\pmb{b})+C(\pmb{-a},\pmb{-b})}{C(\pmb{a},\pmb{b})+C(\pmb{a},\pmb{-b})+C(\pmb{-a},\pmb{b})+C(\pmb{-a},\pmb{-b})},
\end{equation} 
which, in turn, allows calculating $S$ according to Eq. \ref{eq:CHSH}.

\section*{Acknowledgments}
ES, JJ, DO and WT thank Charles Thiel, Neil Sinclair, Morgan Hedges, Thomas Lutz, Khabat Heshami, Marcel.li Grimau Puigibert, Lambert Giner, Andr\'e Croteau and Vladimir Kiselyov, for technical help and/or discussions, and acknowledge funding through Alberta Innovates Technology Futures (AITF) and the National Science and Engineering Research Council of Canada (NSERC). VBV and SWN acknowledge partial funding for detector development from the DARPA Information in a Photon (InPho) program. Part of the research was carried out at the Jet Propulsion Laboratory, California Institute of Technology, under a contract with the National Aeronautics and Space Administration.
\section*{Author contributions}
The SNSPDs were fabricated and tested by VBV, MDS, FM and SWN at NIST and JPL. All measurements were performed by ES and JJ, with help from DO. The manuscript was written by WT and ES, with help from DO. 

\section*{Additional information}
The authors declare that they have no competing financial interests.

Correspondence and requests for materials should be addressed to W. Tittel (email: \mbox{wtittel@ucalgary.ca).}



\pagebreak
\widetext

\begin{center}
\textbf{\large SUPPLEMENTARY INFORMATION}
\end{center}
\setcounter{equation}{0}
\setcounter{figure}{0}
\setcounter{table}{0}
\setcounter{page}{1}
\makeatletter
\renewcommand{\theequation}{S\arabic{equation}}
\renewcommand{\thefigure}{S\arabic{figure}}
\renewcommand{\thetable}{S\arabic{table}}

\section{Superconducting nanowire single photon detector} 
A critical component in our demonstration is the superconducting nanowire single photon detector (SNSPD), which allows for highly efficient detection of the recalled telecom photons with very low dark count rate. Our detector is based on a superconducting tungsten silicide (WSi) nanowire meander and has been developed by some of us at the National Institute for Standards and Technology (NIST) and the Jet Propulsion Laboratory (JPL). Its performance, including high detection efficiency, has been detailed before \cite{Marsili_13S}. In our experiments the detector is mounted on the ADR stage of the cryostat and kept at the same temperature as the fibre memory (0.8--1K). For the detector used for the majority of the measurements \cite{comment1} we found the system detection efficiency to be $\sim$60$\%$, which includes transmission loss through coiled fibres inside the cryostat and splice loss. We measured a dark count rate of around 10~Hz and a dead time of $\sim$35 ns. Furthermore, we found the timing jitter of the detector used for the time-bin projection measurements to be around 250 ps, which allows us to resolve the two, 1.4 ns separated temporal modes that constitute our time-bin qubits. We note that, as opposed to widely used InGaAs-based single photon detectors, our detector does not require any gating signals. Employing the SNSPD in our quantum memory demonstration allows substantially reduced measurement times and increased signal-to-noise ratio.   

\section{The erbium-doped silicate fibre} 
We use a twenty-meter-long, single-mode, erbium doped silica fibre manufactured by INO, Canada. In addition to Er, the fibre contains Ge, P and Al co-dopants. The concentration of the erbium atoms is 80 ppm-wt, leading to 0.6 dB/m absorption at 1532 nm and at room temperature. The fibre is spooled in layers around a copper cylinder with $\sim$4 cm diameter that is thermally contacted with the base plate of an adiabatic demagnetization refrigerator maintained at 0.8-1 K and exposed to a magnetic field of~$\sim$600 G. The fibre is fusion-spliced to standard SMF-28 single mode fibres, resulting in less than 5\% loss per splice. 
 
\section{The AFC memory}
\subsection*{Efficiency}

When triggering a forward propagating echo, as in our investigation, the recall efficiency of an AFC quantum memory, $\eta$, is limited to 54\% \cite{Afzelius_09S}. However, it can reach unity by either satisfying a phase matching condition that leads to re-emission in the backward propagation \cite{Afzelius_09S, Tian_13S}, or by embedding the storage material into an impedance-matched cavity \cite{Afzelius_10aS, Moiseev_10S}. The comb finesse $F$, defined as the ratio of the comb peak spacing to the width of the peaks, the optical depth of the peaks (referred to as $d_1$), and the optical depth of the remaining background (due to imperfect persistent spectral hole burning and referred to as $d_0$) are the main parameters that determine the recall efficiency $\eta$ \cite{Afzelius_09S}. Assuming forward recall, the efficiency is given by:

\begin{equation}
\label{eta}
\eta=\left (\frac{d_1}{F}\right )^2e^{-d_1 /F}e^{-7/F^2}e^{-d_0}
\end{equation}
\noindent
where we assumed Gaussian-shaped comb teeth. At temperatures between 0.8 and 1K, we find $d_0$ to be 1--1.3, which results in around 70\% of the input photons being irreversibly absorbed. Together with a finesse of 2 and $d_1$ of 1, Eq. \ref{eta} yields a recall efficiency of around 1\%, which we confirmed by comparing the number of input and recalled photons. 
 
As will be further described in \cite{Saglamyurek_14bS}, we observe a strong temperature dependence of the hole-burning efficiency, i.e the percentage of the atomic population transferred to other magnetic sub-levels. In addition, spin mixing and stimulated emission during spectral hole burning~\cite{Lauritzen_08S} are expected to further decrease $d_0$ and thus to increase the efficiency of our AFC memory.  Moreover, we observe that the quality of the hole burning significantly depends on the doping concentration of erbium as well as the co-dopants in the fibre. Finally, we note that the strong laser beam used for spectral hole burning propagates in the opposite direction compared to the photons to be stored, which results in reduced pumping efficiency due to imperfect polarization overlap in the non polarization-maintaining fibre. In summary, we expect that a substantial decrease in $d_{0}$, and hence a substantial increase in efficiency can be achieved under optimized experimental conditions.

\subsection*{Storage time}

The storage time in an AFC memory using a two-level atomic system is pre-determined by the spacing $\Delta$ between the teeth of the AFC \cite{Afzelius_09S}:
\begin{equation}
t_\mathrm{storage}=1/\Delta.
\end{equation}
\noindent 
Longer storage times require smaller spacing of the AFC peaks, which is limited by laser jitter and power broadening during spectral hole burning, spectral diffusion \cite{Liu_06S} and, fundamentally, the homogeneous linewidth $\Gamma_\mathrm{hom}$(which is inversely proportional to the coherence time $T_2$),  setting an upper limit to the storage time.

While rare-earth-ion doped glasses have not received as much attention as rare-earth-ion doped crystals, there are several studies of their low-temperature coherence properties (see, e.g., \cite{Huber_84S, Broer_86S, Geva_97S, Mcfarlane_06S, Staudt_06S, Mcfarlane_07S}). For silicate fibres with high erbium doping concentration, coherence times up to 3.8 $\mu s$ have been reported at 0.15 K and 2 T magnetic field \cite{Staudt_06S} --- they are limited by spectral diffusion and coupling to two-level systems (which are specific to amorphous materials such as glass). The application of (relatively high) magnetic fields can suppress these mechanism to some extent. In addition, it has been suggested that reducing the doping concentration may yield longer coherence times due to reduced magnetic interactions between erbium ions \cite{Staudt_06S}. Under our experimental conditions (a magnetic field of $\sim$600 Gauss and a temperature of 0.8--1 K), we have stored photons up to 35 ns using an 8 GHz-wide AFC, yielding a time-bandwidth product of $\sim$300. 

We note that, while the storage time in our investigation was pre-set, there are several, not yet investigated possibilities that may allow deciding the moment of recall after a photon has been absorbed. First, erbium features a large number of hyperfine states, which may allow reversible mapping of optically excited coherence onto spin waves \cite{Afzelius_10bS}. In turn, this would allow selecting the recall time on demand \cite{Afzelius_09S}.  
Second, given the here-demonstrated possibility for persistent spectral hole burning, together with the possibility for reversible broadening using the DC Stark effect \cite{Hastingsimon_06S}, one can turn to the photon echo quantum memory protocol based on controlled reversible inhomogeneous broadening (CRIB)\cite{Kraus_06S,Alexander_06S}. And finally, we emphasize that every application that requires read-out on demand in the temporal domain can be realized using a feed-forward control of the read-out in any other accessible degree of freedom such as in the spectral domain (where, instead of being localized in time bins, photons occupy specific frequency bins) \cite{Sinclair_13S}. Our memory can readily be employed for the latter.

\subsection*{Bandwidth}

The generally large inhomogeneously broadened absorption lines of rare-earth-ion doped solids, in our case around 1.3 THz (equivalent to 10 nm bandwidth), allow tailoring wide-bandwidth AFCs. However, the frequency spacing between the ground state and the shelving level used for persistent hole burning (assumed here to be neighbouring) sets a limit to the bandwidth over which high recall efficiencies can be achieved (see the Supplementary Information of \cite{Saglamyurek_11S}). Hence, the use of shelving levels of magnetic origin is very attractive, in particular in the case of large magnetic sensitivity as in erbium, because they allow for variable and large level-splitting using small magnetic fields. 

However, the magnetic field also affects the persistence of the tailored spectral holes \cite{Hastingsimon_08S, Thiel_10S} --- large fields often lead to reduced persistence. This creates a trade-off between having a long AFC life-time (which benefits the creation of a high-quality AFC through repeated hole burning) and a large splitting (i.e. large bandwidth over which photons can be stored efficiently).

For our AFC preparation, we operate at 0.8 - 1K and apply a magnetic field of 600 Gauss, which results in sufficiently long-lived AFCs. However, as we increase the AFC bandwidth beyond 4 GHz, we see a level-spacing induced increase of $d_0$, which limits our recall efficiency.

\section{Experimental data}

The following tables detail the results of the measurements related to the reconstruction of density matrices (Table. \ref{projections}), and the CHSH Bell inequality tests (Table. \ref{Bell}).

\begin{table}[hh!]  
	\label{projections}
	\begin{tabular}%
		{l c c c c c c c c}
		\hline
		 				     & $\sigma_x\otimes\sigma_x \ $ & $ \ \sigma_x\otimes\sigma_y$ & $\sigma_x\otimes\sigma_z$ & $\sigma_x\otimes\sigma_{-z}$ & $\sigma_y\otimes\sigma_x$ & $\sigma_y\otimes\sigma_y$ & $\sigma_y\otimes\sigma_z$ & $\sigma_y\otimes\sigma_{-z}$ \\
		\hline
		\hline
		$P_\mathrm{in}$ [\%] & $89\pm1$ & $46\pm1$ & $49\pm1$ & $51\pm1$ & $51\pm1$ & $12\pm1$ & $49\pm1$ & $51\pm1$ \\
		$P_\mathrm{out}$ [\%] & $87\pm5$ & $51\pm5$ & $49\pm5$ & $51\pm5$ & $46\pm5$ & $13\pm5$ & $54\pm5$ & $46\pm5$ \\
		\hline
		 				     & $\sigma_z\otimes\sigma_x$ & $\sigma_z\otimes\sigma_y$ & $\sigma_z\otimes\sigma_z$ & $\sigma_z\otimes\sigma_{-z}$ & $\sigma_{-z}\otimes\sigma_x$ & $\sigma_{-z}\otimes\sigma_y$ & $\sigma_{-z}\otimes\sigma_z$ & $\sigma_{-z}\otimes\sigma_{-z}$ \\
		\hline
		\hline
		$P_\mathrm{in}$ [\%] & $49\pm1$ & $49\pm1$ & $89\pm1$ & $11\pm1$ & $51\pm1$ & $51\pm1$ & $13\pm1$ & $87\pm1$ \\
		$P_\mathrm{out}$ [\%] & $47\pm3$ & $49\pm3$ & $87\pm2$ & $13\pm2$ & $53\pm3$ & $51\pm3$ & $12\pm1$ & $88\pm1$ \\
		\hline
	\end{tabular}
	\caption[Joint-detection probabilities for density matrix reconstruction]{\textbf{Joint-detection probabilities for density matrix reconstruction:} Measured joint-detection probabilities for all projection measurements required to calculate the density matrices for the bi-photon state emitted from the source ($\text{P}_{\text{in}}$), and the state after storage and recall of the telecom photon ($\text{P}_{\text{out}}$). Uncertainties indicate one-sigma standard deviations calculated from Poissonian detection statistics.}
\end{table}

 \begin{table}  [hh!]
	\begin{tabular}%
		{l c c c c}
		\hline
						     & $\sigma_y\otimes(\sigma_x+\sigma_y) \ $ & $ \ \sigma_y\otimes(\sigma_x-\sigma_y)$ & $\sigma_x\otimes(\sigma_x+\sigma_y)$ & $\sigma_x\otimes(\sigma_x-\sigma_y)$ \\
		\hline
		\hline
		$E_\mathrm{in}$ [\%] & $55.4\pm1.4$ & $58.8\pm1.3$ & $65.3\pm1.3$ & $58.7\pm1.4$ \\
		$E_\mathrm{out}$ [\%] & $55.4\pm6.2$ & $65.0\pm6.0$ & $57.9\pm5.5$ & $54.9\pm4.9$ \\
		\hline
	\end{tabular}
	\caption{\textbf{Measurement settings and correlation coefficients for Bell-inequality tests.} We measure four correlation coefficients using photon pairs before (in) and after (out) storage of the telecommunication photon. The local measurements are described by superpositions of Pauli operators, and are chosen to maximally violate the CHSH inequality. Uncertainties indicate one-sigma standard deviations and are based on Poissonian detection statistics.}	\
	\label{Bell}
\end{table}

\section*{Calculation of purity, entanglement measures \cite{Plenio_07S} and fidelities}

\noindent Assuming an arbitrary two-qubit input state $\rho$, the \textit{concurrence} is defined as 
\begin{equation}
C(\rho)=\text{max}\left\{0,\lambda_1-\lambda_2-\lambda_3-\lambda_4\right\}
\end{equation}
\noindent where the $\lambda_i$'s are, in decreasing order, the square roots of the eigenvalues of the matrix $\rho (\sigma_y \otimes \sigma_y)\rho^*(\sigma_y\otimes\sigma_y)$ and $\rho^*$ is the element-wise complex conjugate of $\rho$. The \textit{entanglement of formation} is then calculated as
\begin{equation}
E_F(\rho)=H\left(0.5+0.5\sqrt{1-C^{2}(\rho)}\right)
\end{equation}
\noindent where $H(x)=-x\text{log}_2x-(1-x)\text{log}_2(1-x)$. The fidelity between $\rho$ and $\sigma$ is
\begin{equation}
F(\rho,\sigma)=\left(tr\sqrt{\sqrt{\rho}\sigma\sqrt{\rho}}\right)^2 
\end{equation}
\noindent and the \textit{purity} of a state $\rho$ is
\begin{equation}
P=\text{tr}(\rho^2)
\end{equation}
\noindent Finally, the optimum value for $S$ in CHSH Bell-inequality tests (which is the value we expect to measure using optimized measurement settings) can be found from the concurrence:\begin{equation}
S_{th}=2\sqrt{1+C^2}.
\end{equation}

\end{document}